\documentclass[aps,showpacs,preprintnumbers,amsmath,amssymb, 
twocolumn, tightenlines]{revtex4}

\usepackage[dvips]{graphicx}\usepackage{bm}
\usepackage{amsmath}
\usepackage{dcolumn}
\usepackage{bm}

\newcommand{\be}{\begin{eqnarray}}
\newcommand{\ee}{\end{eqnarray}}

\begin{document}
\title{Toward the AdS/CFT gravity dual for High Energy Collisions:
\\ I.Falling into the AdS }

\author{Shu Lin and Edward Shuryak}
\affiliation{ 
Department of Physics and Astronomy,  Stony Brook University , 
Stony Brook NY 11794-3800, USA
}

\date{\today}
\begin{abstract}
     In the context of the AdS/CFT correspondence we discuss the
gravity dual of a high energy collision in a strongly coupled ${\cal
N}=4$ SYM gauge theory. We suggest a setting in which two colliding
objects are made of non-dynamical
heavy quarks and antiquarks, which allows
to treat the process in classical string approximation. Collision
``debris'' consist of closed as well as open strings. 
If the latter have
 ends on two outgoing charges, they are being ``stretched''
along the collision axes.
We discuss motion in AdS of some simple objects first -- massless
and massive particles -- and then focus on open strings. 
We study the latter in a considerable detail,
concluding
that they rapidly become ``rectangular'' in proper time -spatial rapidity
$\tau-y$ coordinates
with well separated fragmentation part and 
a near-free-falling rapidity-independent 
central part. Assuming that in the collisions of ``walls'' of charges multiple stretching
strings are created, we also consider the motion
of a 3d stretching membrane. We then argue that a complete solution can
 be approximated by two different vacuum solutions of Einstein eqns,
with matter membrane separating them. We identify one
of this solution with Janik-Peschanski stretching black hole
solution, and show that all objects approach its (retreating)
horizon in an universal manner.
\end{abstract}

\maketitle
\vspace{0.1in}

\section{Introduction }
The
  AdS/CFT correspondence~\cite{adscft} is a duality of  the conformal
(CFT) $\cal N$=4 supersymmetric Yang-Mills theory
 and string theory in 5d Anti-de-Sitter
space ($AdS_5$). Multiple papers use this fascinating theoretical tool, in
a regime in which the gauge theory is in a strong coupling regime while
string part is in   weak coupling -- the classical SUGRA regime.
The equilibrium finite
temperature version of this correspondence, using a black-hole background, was suggested
by Witten~\cite{witten1}.  Applications of this version of correspondence
to properties of strongly coupled high-T phase of QCD are very actively
pursued: we will briefly review those in the next subsection.

The aims of this series of works are however quite different:
instead of focusing on equilibrium thermal matter, we hope  to develop a
  gravity dual framework to
 time-dependent process of high energy collisions. We will not assume
equilibration or use macroscopic variables like temperature or hydrodynamic
flows:  we hope to 
 be able to understand how they naturally appear for collisions of  large systems.
Instead we focus on motion of strings in $AdS_5$ in this work, and, in the second one,
on ``holograms'' which an observer will see in our world -- the $AdS_5$
boundary -- as a function of time.

Since this is the first paper of the series, we decided to start with rather
extensive introduction, which describes similar works and summaries
our current understanding of the subject.

\subsection{Strongly coupled Quark-Gluon Plasma}
It is well known that  non-perturbative properties of the QCD vacuum phase
-- confinement and chiral symmetry breaking -- are absent
above some critical temperature, where matter is in the so called
Quark-Gluon Plasma (QGP) phase. Although at high $T$ one naturally
expects the QGP to be in a weakly coupled regime,  
 it has been conjectured  recently \cite{SZ12} that 
at least at $T=(1-2)T_c$
-- known as the  RHIC domain -- it 
  is closer to a 'strongly coupled'  regime (sQGP).

This was a significant ``paradigm shift''
in the field, and various  directions toward the understanding
 of sQGP constitute a mainstream of the field. 
Basically there are two competing options: one, based on electric-magnetic
duality \cite{Liao:2006ry}, relates small viscosity and diffusion of sQGP
 to presence of magnetic monopoles and   predicts that it will
disappear at $T$ away from critical region. Another -- based on AdS/CFT --
relates it to ``quasiconformal behavior'' of QGP at $T>2 T_c$.
A comparison between experimental results from RHIC ($T=(1-2)T_c$) with
those at LHC (higher $T$) will hopefully shed light on it in near future.

 Let us only mention 
some important developments related to the latter approach, AdS/CFT.
In a static finite-$T$ setting with AdS-black hole metric 
\cite{witten1}
 the study started with classic results on bulk
thermodynamics \cite{Gubser:1998nz} and transport coefficients~\cite{Policastro:2001yc}:
those works provided intriguing results. It was shown that
while the Equation of State can be quite close to that of weakly coupled
plasma, the transport properties can differ from them by orders of magnitude. It is enough to mention
that while viscosity to entropy ratio is believed to be limited by 
the AdS/CFT value from below \cite{Kovtun:2004de} 
\be {\eta \over s} > {1\over 4\pi} \ee
recent hydrodynamical studies by three groups \cite{visc_hydro}
have concluded that the experimental data on the so called elliptic
flow are better reproduced if this ratio is   even smaller than that!
(For a possible way out of this puzzle, 
also based on AdS/CFT, see e.g. \cite{Lublinsky:2007mm}.)

Then attention focused on high energy jet quenching,
with  the result that a heavy quark pulls a string, with specific
and calculable shape. 
The AdS/CFT  
 result for the drag force  \cite{jet}
and heavy quark diffusion \cite{CT},
turned out to be correctly related by the Einstein relation.
For a recent brief summary see e.g. \cite{Shu_adv06}: it is sufficient 
to mention here that all  these results seem to be in much better
 agreement with  what is seen phenomenologically in
heavy-ion collisions at RHIC than their weak-coupling counterparts.

Further development of the jet quenching problem
was related to the question {\em where does the lost energy go?}.
In a hydrodynamical context it was suggested that
the so called ``conical flow'' \cite{conical} 
 of matter should develop, induced by a heavy charge moving
in a strongly coupled plasma. The ``hologram'' of the dragging string 
has been calculated by
Princeton and Seattle groups \cite{Gubser,Chesler:2007an}: it
described the conical flow picture in  stunning detail.

\subsection{Gravity dual for heavy Ion collisions}
The
  results mentioned above  are all 
equilibrium ones, obtained using static AdS-black hole
metric.  Although they should be applicable
for a macroscopically large and slowly expanding
fireball,  one may proceed to more
demanding issues related with  AdS/CFT in time-dependent
out-of-equilibrium
 setting. Those will  provide new insights into 
equilibration issues, explaining when
and with what accuracy  thermo- and
  hydrodynamics
become applicable.

In AdS/CFT language going from cold vacuum to hot plasma
means going from pure AdS (extremal black hole) to black-hole AdS
via creation of trapped surface. Therefore the problem to
be considered is a kind of gravitational collapse, occuring
in gravity-dual as a result of high energy collision. 

The quest for black hole formation in collisions has a long history
we would not attempt to review here. Let us just mention that
it was discussed for ``real'' gravity at colliders, which
may get possible provided it gets strong due to extra dimensions. 
Black hole production in 
AdS spaces were discussed both in cosmological brane world
models, as well as in AdS/CFT framework
from
late 90s: we only mention few papers most related to our work.
 Black hole emerging from collisions were
discussed in $AdS_5$ background by Horowitz and
Itzhaki\cite{Horowitz:1999gf}, who considered
departing black hole. 
 Giddings and Katz\cite{Giddings:2000ay}
have discussed holograms of the falling objects
in $AdS_5$ background, in cosmological setting
(which has some differences with AdS/CFT one in
boundary conditions).
 In \cite{Matschull:1998rv} a solution for black hole creation from
collision of particles was obtained for a simpler case of
 $AdS_3$ background. It was recently further
 stuided by Kajantie et al.
\cite{Kajantie:2007bn}.

 In the context of gravity dual to heavy ion collisions, the
problem of black hole formation was discussed by 
Sin, Shuryak and Zahed \cite{SSZ} (SSZ below).
 One specific solution they discussed in the paper 
was a ``hologram'' of a departing black hole,
corresponding to a spherically symmetric (Big-Bang-like) solution
 with a decreasing
$T$ . SSZ also proposed two other idealized settings, 
with d-dimensional stretching, corresponding for d=1 to
a collision of two infinite  thin walls and subsequent ``Bjorken''
rapidity-independent expansion\cite{Bj}, with 
 2d and 3d corresponding to cylindrical and spherical 
relativistic collapsing walls.

Janik and Peschanski\cite{Janik2} (below referred to as JP) have 
addressed
the simplest wall-on-wall collision.
In this case the time and longitudinal coordinate
$x_1$ are naturally substituted by  
the proper time and spatial rapidity
\be \tau=\sqrt{t^2-x_1^2},\,\,\, y={1\over 2}log({t-x_1\over t+x_1})
\label{eqn_tautheta} \ee
 since the rapidity-independent solution
  depends on only $\tau$.
Instead of solving Einstein equations with certain
source, describing
gravitationally collapsing ``debries'' of the collision,
JP applied an ``inverse logic'',  extrapolating
  into the bulk the metric which yield
 expected hydrodynamical solution at the boundary.
JP found an $asymptotic$
(large-time) solution for a 
 ``stretching AdS-BH''. 
As expected, it indeed possesses a horizon moving away from
the $AdS$ boundary, as $z_{horizon}\sim \tau^{1/3}$.
 A very 
 important feature of the leading-order  JP solution
is entropy conservation:
is that while their presumed horizon is stretching in one 
direction and contracting in others, to the {\em leading order}
two effects compensate each other and keep the
 {\em total horizon  area constant}. 
We will discuss a bit more this solution and use it in section \ref{sec_jp}.

Further discussion of the subleading 
(next power of inverse time) terms 
has been made by Sin and Nakamura  \cite{SinNak}  who 
 identified corrections to the JP solution  with the viscosity effects.
 Terms of still higher order have been subsequently studied
 \cite{Heller:2007qt}, but eventually
Janik et al  \cite{Benincasa:2007tp} concluded that
the expansion series are inconsistent beyond the first few orders.
 Our view is that this is how it should be, and
the arising near-horizon singularity indicate that
presence of matter term (absent in JP) 
is inevitable.


 Unlike JP et al we will not
 use any
  ``inverse logic'' and will not be  looking for the 
solutions corresponding to pre-determined
hydro on the boundary. Instead we will focus on
 the formation stage, whether black hole is or is not formed,
 and will  $calculate$ the (time-dependent) stress tensor
 on the boundary, whether it is a hydro-type on not. 
 
\subsection{Hadron collisions in QCD, the Lund model and the ``Color Glass''}

Rather early in development of QCD, when the notion of confinement and
electric flux tubes -- known also as the QCD strings -- were invented in 1970's,
B.Andersen and collaborators \cite{Andersson:1983ia} developed what gets to be known as the Lund model of hadronic collisions. Its main idea is that during short time
of passage of one hadron through another, the strings can get reconnected,
and therefore with certain probability some strings become connected to
color charges in two different hadrons. Those strings get stretched longitudinally
and then break up into parts, making secondary 
mesons and (with smaller probability)
baryons. Many variants of string-based models were developed,
 and some descendants --e.g. PYTHIA -- remain
 widely ``event generators'' till today.

 If there are several string stretched, it is usually assumed that
 both their interaction and influence on breaking is negligible.
 
 However if one either considers very high energy collisions, when a single
 hadron should be viewed as being made of many color charges (partons),
 or heavy ion collision, a different asymptotic picture has been proposed.
 McLerran and Venugopalan \cite{McLerran:1993ni} argued that instead of multiple string
 the fields produced should be considered as classical gauge fields
--known as Color Glass model -- and their
 subsequent evolution be derived from solution of classical Yang-Mills
 equation \cite{RAJU}. 
 They suggested this regime is true at very high parton
density, when the effective coupling is weak.
  Accepting the Color Glass
  picture as a correct asymptotic  for very high parton density
 and large saturation scale $Q_s\rightarrow \infty$, one still wanders what should happen
 in the case of intermediate scale
  $Q_s\sim .3-1.5 \, GeV$.

   Recent developments of the so called AdS/QCD proposed a view
 that this interval of scales
in QCD constitute a  ``strong coupling window''. 
 In particular, Brodsky and Teramond \cite{Brodsky:2007vk} 
have argued that
 the power  scaling observed for large  number of exclusive processes
 is not due to perturbative QCD (as suggested originally in 1970's) but
 to a  strong coupling regime with near-constant coupling
(quasi-conformal regime).  Polchinski and Strassler
 \cite{Polchinski:2001tt} have shown that in spite of exponential
string amplitudes one does get power laws scaling for exclusive
processes, due to convolution (integration over the $z$ variable)
with the power tails of hadronic wave functions.
 One of us
 proposed a  scenario \cite{mydomain} for AdS/QCD
 in which there are two domains, with weak and strong coupling. The
 gauge coupling rapidly rises at the ``domain wall''
associated with instantons. 
 Such approach looks now natural in comparison to what happens in
 heavy ion/finite T QCD, where we do know  that at comparable parton
 densities  the system indeed is in a strong coupling regime. 

 \subsection{The goals of this series of papers}
 In short, 
it is to study self-consistently the collision process in AdS/CFT. 
 For hadronic collisions we basically follow
QCD-string-inspired (Lund) picture of the
 collision. While QCD phenomenology focused on ``string breaking'',
 in AdS/CFT setting we will have instead their ``falling'' (departure
 from the boundary) into the IR. 

  In this paper
 we will study in detail motion of ``debris'' -- massless and massive particles
 and open strings, and membranes
 -- in $AdS_5$. 
  In the second paper we will calculate the corresponding
 ``holograms''
of these objects -- 
 the stress tensor of matter created on the boundary.
Although ``debris'' fly away into the 5-th direction,
the usual energy and momenta are conserved in our world,
and those ``holograms'' describe a flow of matter outward from
the collision point. As we mentioned already, this can be viewed
as a strongly-coupled version of Color Glass, put in the
realm of $\cal N$=4 SYM theory.

 We hope in subsequent works to
go beyond the linearized gravity and 
follow nonlinear effects leading to
  a gravitational collapse of debris and formation
  of trapped surfaces. This would be dual to information loss (entropy production)
  and appearance of equilibration.

\begin{figure}[t]
\centerline{\includegraphics[width=8cm]{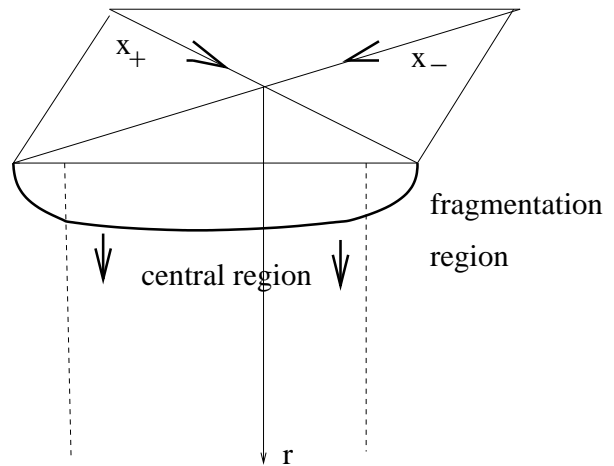}}
\caption{\small Schematic view of the collision setting.
The classical heavy charges move along directions $x_\pm$ and collide
at
the origin. String snapping leads to longitudinally
stretched strings (wide black line) which are also
extended into the 5-th coordinate $r$ toward the
AdS center at $r=0$. The heavy charges move on the plane $r=\infty$
} \label{fig_coll_q}
\end{figure}
\section{The setting}

  One important suggestion made by SSZ is that heavy ion collisions
posses ``some internal high 
momentum scale'', usually called $Q_{saturation}$, related to
high density of color charges in boosted heavy ions.
In order  to model it more simply,  we  now propose
substitute
energetic light quarks  by heavy ones,
 with the mass $M_Q$ of  
heavy fundamental quarks $Q$ introduced into AdS/CFT
via $D_7$ brane\cite{Karch:2002sh}. As soon as  $M_Q$ is at the scale of
$Q_{saturation}$, 
it  makes little dynamical difference: but in the AdS/CFT
language treatment of heavy quarks is simpler, as they are
sources of classical strings. (This simplifying 
feature has been put to heavy use
in  treatment of the heavy quark jet quenching  \cite{jet}.)

We will further assume that heavy quarks 
 have no dynamics of their own, as they are moving along  straight
lines 
\be x_{\pm}=x_1\pm v t\ee
with  constant velocity $v$, both before and after the collisions,
see Fig.\ref{fig_coll_q}. If so, there is no conventional
gluonic radiation on the brane or gravitational radiation
from them in the bulk, as there is no acceleration.

The dynamical objects we will focus on   are  
classical strings, ending at these heavy quarks
and propagating in the bulk (for
 metrics changing from  AdS to JP-like one).
We will study which solutions exist as a function of collision
rapidity and whether 
they are stable or not: we will conclude that
at sufficiently large $v>v_c$ these strings basically
go into free fall toward the AdS center. 

 The next step is to consider not a single pair
of charges (a single stretching string), but many. 
One limit is a pair of colliding ``walls of matter'',
containing multiple heavy quarks. For 
simplicity, think of these two walls
as CP mirror images of each other,  
 made of colorless ``dipoles''. ``Snapping'' of their string
at the collision leads to multiple strings, all of which being
 stretched longitudinally.

 We then
argue that many such strings combined could be considered 
as a thin singular  sheet of matter,
referred to below as ``membrane''. 
(Note an important distinction between a membrane and
a ``true brane'': since
 the former has only energy-momentum but
 lacks the RR charges and 
consequent Coulomb repulsion, it cannot ``levitate'' like
branes, and simply falls under gravity.)

 It has been shown by Israel \cite{Israel} 
how a gravitational collapse of a thin layer
of matter can be described via  two different
discontinuous $vacuum$ solutions of the Einstein
equation without matter ($T_{\mu\nu}=0$).  Self-consistency
of the solution
is then reached by fulfilling covariant {\em junction conditions}, 
resulting in membrane equation of motion. 

\begin{figure}[t]
\centerline{\includegraphics[width=8cm]{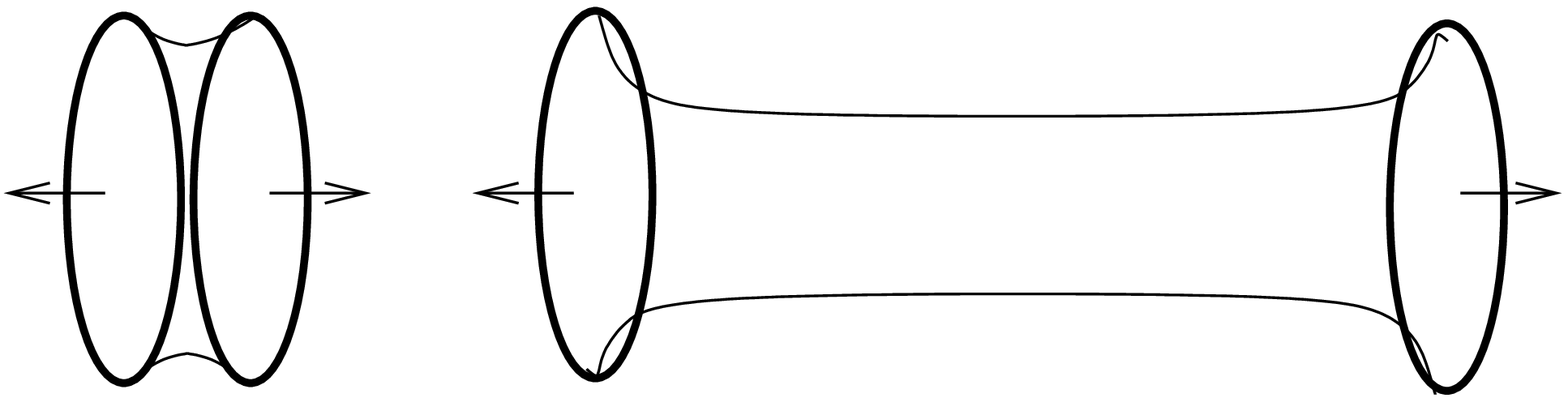}}
\vskip .3cm
\centerline{\includegraphics[width=8cm]{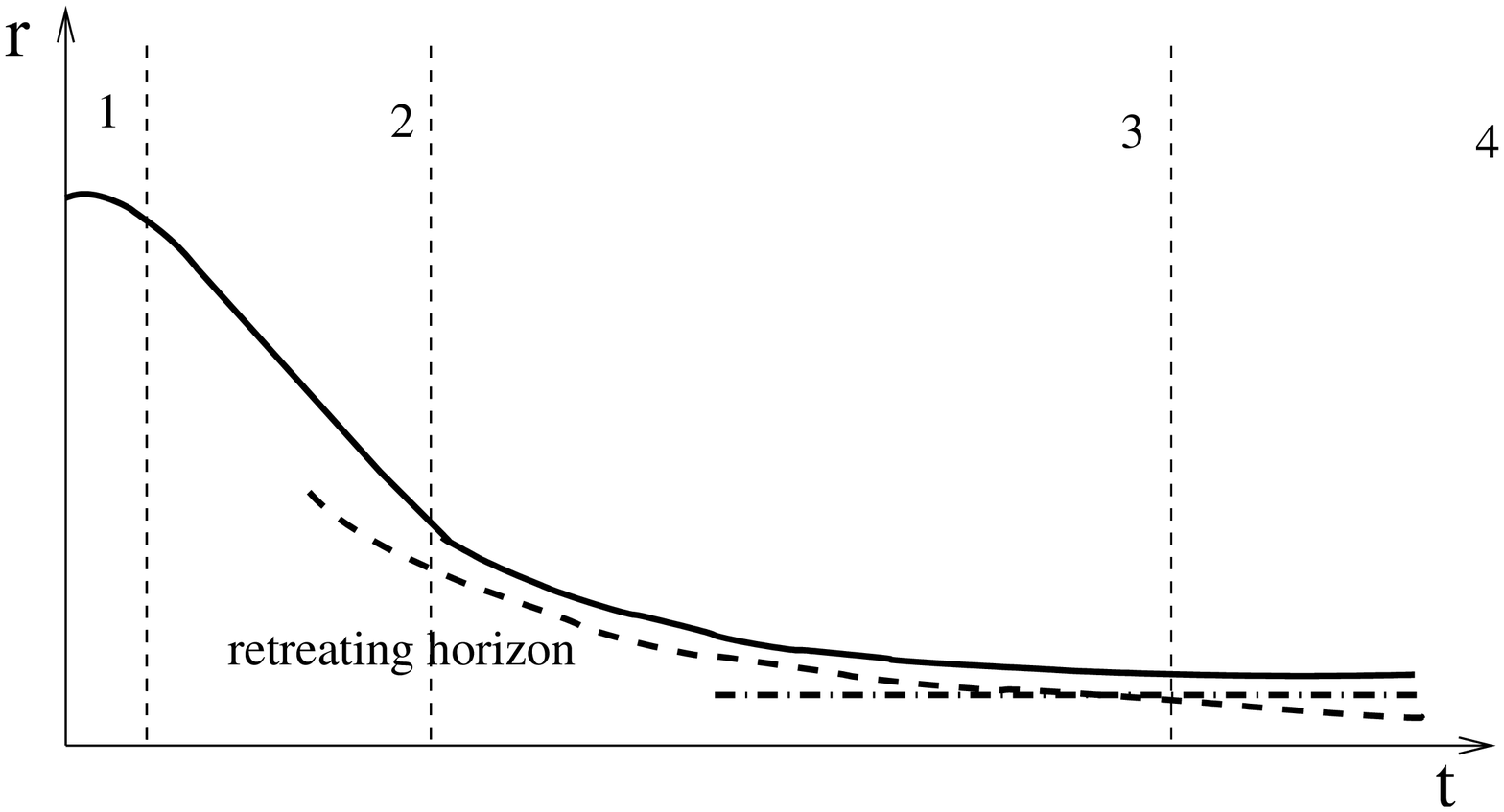}}
\caption{\small 
(a=upper) Two snapshot of the membrane shapes, at different time
  moments. See text for explanation of the coordinates. 
(b=lower)Schematic view of the four periods in
gravity dual solution in which falling objects are
(1) accelerated into the 5-th dimension $r$ till they reach
a relativistic velocity $v\approx 1$, then (2) continue their
 relativistic
fall till  (3) breaking near the retreating horizon.
} \label{fig_phases}
\end{figure}

The issue of self-consistency will not be addressed in this work:
we will discuss below falling of various objects -- particles and open
strings, as well as 3+1 membranes -- ignoring for now
the effect of their own weight on the metric. 
The proposed evolution of the system is explained schematically
in Fig\ref{fig_phases}. Part (a) of it shows some snapshots
of this surface, at some early time and then at a later stage.
The horizontal direction is the collision direction
$x_1$ while the one along the circles represent any of the two other
transverse directions $x_2,x_3$ (on which no dependence
is expected). The radial direction $r$ in part (b) is the
5-th AdS radial direction, a distance from the AdS center.
Since the ``membrane'' is  being stretched in  $x_1$
(linearly in time),  it has to retreat in $r$ and become
a thinner cylinder, just as a stretching 
soap film will do in a similar setting.  

At this point we would like to emphasize a close analogy,
as well as differences, with the jet quenching problem.
%
 One studied first 
a single falling string
governed by simple Nambu-Goto action and the overall metric. 
The complicated picture of matter flow is then 
 recovered   using weak (linearized) gravity.
One difference is that in a jet quenching
problem the string is stationary (in the charge frame)
while in our case it is not. Furthermore, we will discuss also
 multiple strings, which may form another singular object -- the
$membrane$. Also the metric in our problem
is first considered to be just AdS, but
  eventually it will be
non-trivially affected by the membrane's own weight. 
If so, one should no longer use the linearized gravity
 but  solve Einstein equations in its full
nonlinear form. 

Needless to say, this is a very difficult task, amenable to
analytic treatment only if some drastic simplifications are made.
A scenario outlined in Fig\ref{fig_phases}(a) would have
metric dependent on 3 variables: time, longitudinal direction
and the AdS radial one, $t,x_1,r$. We thus propose a further 
simplification of the problem: changing variables to
proper time and spatial rapidity (\ref{eqn_tautheta})
we would look for $y$-independent solutions,
corresponding to purely cylindrical part of the membrane
in the middle of Fig\ref{fig_phases}(a),
ignoring the curved ``fragmentation'' regions.
 With only two variables,
$\tau,r$ one 
has a problem of similar  level of complexity as the one addressed
by
Israel\footnote{Except that in Israel's problem of
non-stretching black hole the horizon is
stationary, while in our case it is moving. }, for a spherical gravitational 
collapse.

Further clarification of the  proposed scenario is shown in
Fig.\ref{fig_phases}(b), displaying a
trajectory of the membrane $r(\tau)$.
 During the first stage of the process the ``debris'' of a collision
in a bulk
 -- the particles and open strings -- are
 accelerated by the AdS gravity
and fall into the 5-th dimension  till they reach
the relativistic velocity $v\approx 1$ (stage 2)). If there be
only one object falling, its gravity being negligible
compared to overall gravity of the $N$ branes at the AdS center
and they would simply continue their
 relativistic
fall. However large number of them have enough
mass to create a horizon which suddenly slows down
the membrane (as a distance observer sees it\footnote{  
As usual for a gravitational collapse, in a co-moving  
frame
the horizon is not important and is crossed, which is not important
for us to follow in this work.}
): at
 stage 3 the membrane is trailing the receding horizon
(the dashed line).

 If we would discuss pure AdS/CFT theory this would be
the end of the story: but in other more QCD-like
setting one can have an additional potential
which will stop membrane because of existence
of a stationary ``deconfinement'' horizon. If so,
the system reaches a ``mixed phase'' era with 
stationary horizon and fixed $T$, 
similar to static fireball discussed by Aharoni et al \cite{Aharoni} 
except that in our setting the
longitudinal stretching continues.

 The trajectory of the
collapsing matter sheet should be such as to provide
 a consistent solution to Einstein equations,
combining the JP-like vacuum solution outside the falling sheet,
with the ``stretching AdS'' inside it.

The paper is structured as follows. In the next section we
solve equation of motion for different objects
falling in AdS. We start
with massless and massive particles in subsection 
\ref{sec_stone_ads}.

The main part of this work is study of
the open strings, being stretched between two departing charges.
We derive analytically the so called $scaling$
 (factorisable) solution in section \ref{sec_scaling}.
 Similar solutions have been used previously
in connection to anomalous dimensions of ``kinks''. 
 New part is discussion of the limits for its existence
and stability.

We then find more general non-factorisable solutions
in section \ref{sec_strings_ads} which can only be obtained
numerically. We  find that in proper time -spatial rapidity
coordinates $\tau,y$ we use those basically
becomes ``rectangular'' ,
with a nearly free-falling rapidity-independent part. We conclude this
section with results for 
 falling membranes. The next section starts with
an introduction to the issue of ``stretching black holes''
in section \ref{sec_jp}, and concludes with section
\ref{sec_objects_JP} in which we show that all objects considered
above are approaching the (retreating) horizon in a very universal
fashion. We conclude with some discussion and outlook in section
\ref{sec_summary}.

In the second paper of the series we will calculate back reaction of
gravity, by solving linearized Einstein equations and obtaining 
 stress tensor on the boundary (``holograms'') for some of these
 falling objects.

\section{Objects falling in $AdS_5$}

The collisions creates a lot of ``debris'' in form of various
excitations. Since we would like to follow the collision
in the bulk, we naturally have to think of them in terms of string
theory. Thus there are the following types of objects:
(i) massless and massive particles;  (ii) open strings, with ends at the
receding walls; (iii) membrane.
 The ``open string'' category is naturally split into
``mesons'' with both ends on the $same$ wall, and 
``stretched strings'',
with both ends attached to different walls and 
moving in the opposite direction.
We will consider a set of multiple strings copied many times
in transverse dimensions $x_2,x_3$ as a 3-d membrane. The validity
of this approximation will be explained later.

\subsection{Falling particles}
\label{sec_stone_ads}
As is usually done in this kind of problems, the
AdS radius is inverted, so that a coordinate $z=1/r$ is used
instead of $r$. The AdS boundary is thus at $z=0$ and ``falling''
objects move away from it toward infinity.
The $AdS_5\times S^5$
 metric in such coordinates is
\be ds^2={R^2 \over z^2}(d\vec x^2-dt^2+dz^2)+R^2\,d\Omega_5^2
\ee
where the last term, related to angles of $S^5$ is of no importance
in this work. We choose to work in $\tau$, $y$ coordinates mentioned above
(\ref{eqn_tautheta}). 
The metric is translated into the following form:

\be
ds^2={R^2 \over z^2}(-d\tau^2+\tau^2dy^2+dz^2)
\label{eqn_AdS_5_g}
\ee
where we ignore the transverse coordinates and the $S_5$ part.

One feature of $AdS_5$ metric is its boost invariance, the importance of
which will be seen later.
Let us assume particles move with constant
spatial rapidity $y$, so the trajectory can be described by $z(\tau)$.
Massless particles move along the geodesics with zero interval
$ds^2=0$ which 
in the metric (\ref{eqn_AdS_5_g}) simply means $z=\tau$. 

Massive falling objects were already discussed in \cite{Danielsson:1998wt},
but here we present it in a different form, more
closely resembling much more nontrivial ones in the next sections.
Using the coordinate time $\tau$ one simply write down
the interval as an action for a particle moving in the
5-th direction 
of
\be S \sim \int d\tau {\sqrt{1-\dot z(\tau)^2} \over z(\tau)} \ee
where the non-trivial trace of the AdS metric is $z$
in the denominator. This
 leads to well known EOM
\be \ddot z(\tau)= {1-\dot z(\tau)^2 \over z(\tau)}
\label{eqn-stone}
\ee
 Nonrelativistically, one can neglect $\dot z(\tau)$ and 
think thus about a motion in a logarithmic potential well\footnote{The
reader may ask why we don't refer to conserved energy, which will make
this much simpler: the reason is the next section would not
have this avenue open for us.}.
Ultrarelativistically, one finds instead that as $\dot
z(\tau)\rightarrow 1$ the acceleration goes to zero, as needed.
Thus, in the standard coordinates, very little seems to happen
after the particle reaches ultrarelativistic regime: it runs forever
toward $z\rightarrow\infty$ with  speed of light. But this is a
(well known) illusion due to relativistic time slowing:
in its own proper time, the particle 
continue to accelerate and reaches the AdS center
in finite proper time. 

 This EOM is easily integrated yielding 
\be z(\tau)= \sqrt{\tau^2+v_0z_0\tau+z_0^2} \ee
%



\subsection{Falling open strings: the scaling solution}
\label{sec_scaling}
After this little warm-up, let us consider motion of the 
open strings. Its action is given by Nambu-Goto, and if one 
ignores two transverse coordinates $x_2,x_3$ and
uses as two internal coordinates 
 the    $t,x$ (time and longitudinal coordinate)
 the string is described 
by by one function of two variables $z(x,t)$. The corresponding string
action is then
\be S=-{R^2\over 2\pi\alpha'}\int dt\int {dx\over z^2}
 \sqrt{1+({\partial z\over \partial x})^2-({\partial z\over \partial t})^2}\ee
Note that only one term,
the time derivative, is different from long-used static
 action
used in \cite{MALDA2_REY} for static calculation of the inter-charge
potential.
The boundary conditions would be $z=0$ at two rays $x=\pm v t$,
the world lines of the heavy quarks.(The boost invariance
of the $AdS_5$ metric allows us to work in a frame where the open string
endpoints move with opposite velocities)

Translating into the $\tau,y$ language, the boundary conditions are
now determined at fixed $y=\pm Y$ where $v=tanh Y$ and $Y$
is the rapidity of the heavy quarks (colliding walls).
 by doing so, we transfer time dependence from the boundary
conditions into the equations themselves. 
The corresponding action is now
\be 
S=-{R^2\over 2\pi\alpha'}\int  { \tau d\tau dy \over z^2}
 \sqrt{1- \left( {\frac {\partial z}{\partial \tau}} \right) ^{2}
+{\frac { \left( {\frac {\partial z}{\partial 
y}} \right)^{2}}{{\tau}^{2}}}
}
\ee

  Before solving the corresponding equation in full, we will
 first discuss ``scaling'' solutions
in
the separable form 
\be z(\tau,y)={\tau\over f(y)}\ee
suggested by conformal properties of the theory. 
Such solutions were known
in literature \cite{Gross_Makeenko},
 in Euclidean context,
they were used for AdS/CFT calculation
of the anomalous dimensions of ``kinks'' on the Wilson lines
(of which our produced pair of charges is one).

 The scaling ansatz leads to a simple 
action
\be 
\label{eq:area}
S=-{R^2\over 2\pi\alpha'}\int  { d\tau dy \over \tau} \sqrt{f'^2
+f^4-f^2} 
\ee
 Using the fact that $y$ does not
appear in the action, there is a conserved ``energy''
\be {V \over \sqrt{f'^2+V}}=E \ee
with the ``potential'' $V=f^4-f^2$, and thus the derivative
of the function $f$ can be readily obtained
\be f'={\sqrt{V(V-E^2)}\over E} \label{eq:df}\ee
Note that the function
$f$ decreases from infinity on the boundaries
 to its lowest value at the middle of the string which we will call $f_0$,
so $f>f_0$. At $f=f_0$ the derivative 
 vanishes, so (\ref{eq:df}) provides also a simple equation  
$f_0^4-f_0^2-E^2=0$ relating $E$ to $f_0$. 

\begin{figure}[t]
\centerline{\includegraphics[width=8cm]{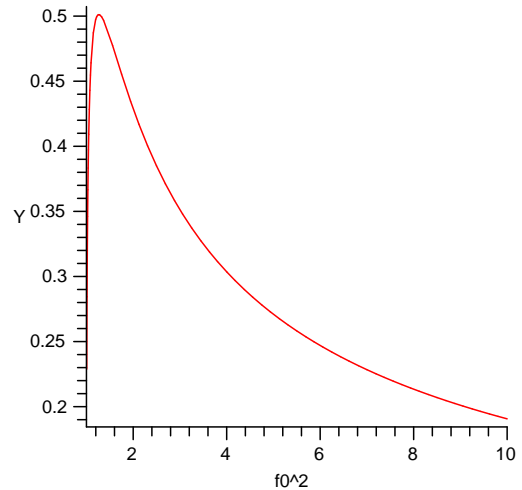}}
\caption{\small Rapidity of the collision $Y=arctanh(v)$ 
vs $f_0^2$. The maximum gives
a critical rapidity $Y_c$. For $Y<Y_c$, two $f_0^2$ are possible, corresponding
two string configurations. For $Y=Y_c$ ,only one $f_0^2$ is possible.
The region $Y>Y_c$ cannot be reached
} \label{ff0_y}
\end{figure}

Integration of (\ref{eq:df}) gives the following solution
\be
\label{eq:sol}
&&y=f_0\sqrt{\frac{f_0^2-1}{2f_0^2-1}} {\it F} 
\left( \sqrt {{\frac {{f}^{2}-{f_0^2}}{{f}^{2}-1}}},
\frac{f_0}{\sqrt{2f_0^2-1}} \right) \nonumber\\
&&-\frac{1}{f_0}\sqrt{\frac{\left( f_0^2-1 \right)^3}{\left(2f_0^2-1 \right)}}
{\it \Pi} \left( \sqrt {{\frac {{f}^{2
}-{f_0^2}}{{f}^{2}-1}}},{1\over{f_0^2}},\frac{f_0}{\sqrt{2f_0^2-1}} \right)
\ee
where ${\it F}$ and ${\it \Pi}$ are elliptic integral of the first and 
the third kind. $f_0^2$ depends on collision rapidity  $Y=arctanh(v)$
 via the boundary condition
at $f(Y)=\infty$, as shown in  Fig.~\ref{ff0_y}.

The existence of a 
maximum 
means that there are no scaling solutions when the rapidity $Y$
is larger than some  
critical value, while if the quarks move on the 
boundary slower that the critical rapidity,
 there are $two$  solutions.

  In order to characterize the solutions, it is useful to
introduce
``effective potential'' for two separating quarks for each scaling
solution, defined as instantaneous energy
 $U=\Delta S/\Delta t$,
 where $\Delta S$ is action given by the area 
of the string world sheet, $\Delta t$ is the time interval.
$U$ needs to be regulated,
which is obtained by subtracting the Wilson loop corresponding to 
two non-interacting moving quarks. In other words, we calculate the subtracted
 area:

\be
&&S_{reg}=-{R^2\over 2\pi\alpha'}\int {dt\over t}\int dy \sqrt{f'^2+V}-\int_0^\infty df \nonumber \\
&&=-{R^2\over 2\pi\alpha'}\int {dt\over t}
\left({\int_{f_0}^\infty df \sqrt {{\frac {V}{V-E^2}}}}-{\int_0^\infty df} \right)
\ee

\begin{figure}[t]
\centerline{\includegraphics[width=8cm]{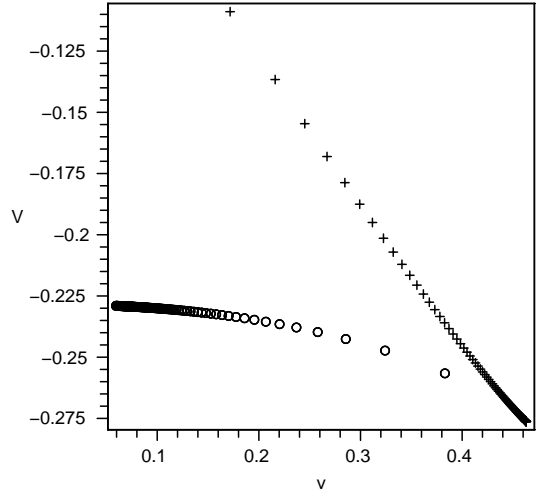}}
\caption{\small the potential V as a function of v for different branches
 of solution. circles for large-$f_0^2$ branch, crosses for small-$f_0^2$ branch
V is plotted in unit of $\sqrt{g^2\,N}/L$
The potential from the large $f_0^2$ branch is lower than that from 
small $f_0^2$ branch
} \label{potential}
\end{figure}

The second term corresponds to $f'=\infty$, precisely the 
straight string going in $z$ direction, which is AdS solution for  
 a moving quark.
Note that we have
 switched to $t$,$y$ coordinates, which does not change the form 
of the string action (\ref{eq:area}). 
With this prescription, we calculated
$U$ for solutions in both branches, which are compared
in Fig.~\ref{potential}. 
 The solution with the lower potential has a chance
to be the stable one, while the higher potential one
(with large $f_0$, or longer string) must be
metastable. 

  Let us now comment on the $small$ v limit of the scaling solution.
At large separation (realized at late time)  the quarks
 can be considered as quasi-static.
At small v, or large $f_0^2$, the effective potential
can be simplified to the following form
\be
\label{eq:nS}
&&dS_{reg}/dt=-{R^2\over 2\pi\alpha'}\int df\left( 
\sqrt{ V \over V-{E}^{2}}-1 \right)/t \nonumber\\
&&=-{R^2\over 2\pi\alpha'}
\left(- 0.5991\,\sqrt{f_0}- 0.1780\,
{1\over {f_0}}\right)\,{2v \over L}
\ee
and relate more simply the velocity and $f_0$
\be
\label{eq:v}
v={0.5991\over{f_0}}- {0.03115\over{f_0}^{3}}
\ee

Combining (\ref{eq:nS}) and (\ref{eq:v}), we obtain
the effective potential for small velocity
and large separation to be

\be
V=0.2285\,\frac { \left( 1+ 0.6830\,v^2 \right) \sqrt {g^2\,N}}{L}
\ee
The coefficient in front (the
leading term at $v\rightarrow 0$) coincides with
the well known coefficient of static Maldacena potential.

The second term is thus the velocity-dependent ``Ampere's law''
 $O(v^2)$ correction to it. We are not aware of any other
previous calculation of this term, except for the paper
by Zahed and one of us \cite{SZ_spin} in which, based on resummation
of ladder diagrams via Bethe-Salpeter eqn, the result was
that the velocity dependence is
\be U(v)/U(v=0)=\sqrt{1-\vec v_1 \vec v_2}\approx 1+.5 v^2+... \ee
It is close but not  the same\footnote{The situation in which two
 charges move in the same direction is just a Lorentz boosted static
 solution: in this case a square root of $v$ in the Lorentz factor
is of course obvious.}.

Both branches of the
 scaling solution was also confirmed by solving the 
equation numerically, starting from the middle point and
scanning all values of
$f_0$.

  The applicability of the scaling solution for a particular $Y$
depends of course not only on availability of a solution, but also
on its $stability$ i.e. how does the scaling solution
evolve with time($\tau$), given some perturbation at initial time.
Denoting scaling solution $g_s(y)={1\over{f(y)}}$ and perturbation as

\be
\label{eq:pt}
z(\tau,y)=\tau g(\tau,y) \hspace{1cm}
g(\tau,y)=g_s(y)+\delta g(\tau,y)
\ee
we want to know whether the perturbation will grow or decay with time.
The EOM for $g(\tau,y)$
\be
\label{eq:nonscaling}
&&-2- {\tau}^{3}g ({\frac {\partial g}{\partial \tau}}) ^{3}
 +2\tau g ({\frac {\partial g}{\partial y}})^{2}{\frac {\partial g}{\partial \tau}}
-2\tau g ^{2} {\frac {\partial g}{\partial y}}{\frac {\partial ^{2}g}{
\partial \tau\partial y}}
+ \nonumber\\
&&{\tau}^{2} g ({\frac {\partial g}{\partial y}})^{2}{\frac {\partial ^{2
}g}{\partial {\tau}^{2}}}
+2\tau {\frac {\partial ^{2}g}{\partial {y}^{2}}} g ^{2}
{\frac {\partial g}{\partial \tau}}
 + {\tau}^{2}{\frac {\partial ^{2}g}{\partial {y}^{2}}} g 
({\frac {\partial g}{\partial \tau}})^{2}-2{\frac {\partial g}{\partial y}}
- g^{4} \nonumber\\
&&- {\frac {\partial ^{2}g}{\partial {y}^{2}}} g^{2}
 +7  \tau g{\frac {\partial g}{\partial \tau}}
-3\tau g^{3}{\frac {\partial g}{
\partial \tau}} -3{\tau}^{2}  g ^{2} ({\frac {\partial g}{
\partial \tau}}) ^{2}+{\tau}^{2}g
 {\frac {\partial ^{2}g}{\partial {\tau}^{2}
}} \nonumber\\
&&+3g ^{2}+2{\tau}^{2} ({\frac {\partial g}{
\partial \tau}})^{2}
+ {\frac {\partial ^{2}g}{\partial {y}^{2}}} g ^{3}
-2{\tau}^{2} g {\frac {\partial g}{\partial y}} {
\frac {\partial g}{\partial \tau}}{\frac {\partial ^{2}g}{\partial \tau\partial y}}=0
\ee
can be used by
plugging (\ref{eq:pt}) in (\ref{eq:nonscaling}), and keeping only term linear
in $\delta g(\tau,y)$(consider only sufficient small perturbation), we obtain
the following linearized EOM for the perturbation:

\be
&&\left[ A+B\frac{\partial}{\partial\tau}+C\frac{\partial}{\partial y}+
D\frac{\partial^2}{\partial\tau\partial y}+E\frac{\partial^2}{\partial\tau^2}
+F\frac{\partial^2}{\partial y^2} \right] \delta g(\tau,y)\nonumber\\
&&=0
\ee

with
\be
&&A=g_s'' g_s^2+6g_s-4g_s^3-g_s''\nonumber\\
&&B=\tau(2g_s g_s'^2+2g_s'' g_s^2+7g_s-3g_s^3)\nonumber\\
&&C=-4g_s'\nonumber\\
&&D=-\tau(2g_s^2 g_s')\nonumber\\
&&E=\tau^2(g_s g_s'^2+g_s)\nonumber\\
&&F=g_s^3-g_s\nonumber\\
\ee

define $\tilde\tau=ln\tau$ as our time, the EOM simplifies to:

\be
\label{eq:varg}
&&\left[ \tilde A+\tilde B\frac{\partial}{\partial\tilde\tau}+\tilde C\frac{\partial}{\partial y}+
\tilde D\frac{\partial^2}{\partial\tilde\tau\partial y}+\tilde E\frac{\partial^2}{\partial\tilde\tau^2}
+\tilde F\frac{\partial^2}{\partial y^2} \right] \delta g(\tau,y)\nonumber\\
&&=0
\ee

with
\be
\tilde A=A, \tilde B=B-E, \tilde C=C, \tilde D=D, \tilde E=E, \tilde F=F
\ee
(To make it easier to get all these functions
 one can approximate scaling solution
  $g_s(y)$ with
some parameterizations: we found that
$({\frac{g_s}{g_s(0)}})^3+({\frac{y}{Y}})^n=1$ fits
all the scaling solution 
very well.)

We need to seek eigenfunction 
$\delta g(\tau,y)=e^{\lambda\tilde\tau}\psi(y)$ satisfying (\ref{eq:varg}) and
boundary condition $\psi(y=\pm Y)=0$
In general, out of many eigenvalues $\lambda$ we should be interested
in those with positive real part, which will allow us
to conclude when the solution is unstable.

The eigenfunction results in the following EOM:
\be
\left[ C_0+C_1\frac{\partial}{\partial y}+C_2\frac{\partial^2}{\partial y^2}
\right]\psi(y)=0
\ee

with
\be
&&C_0=\lambda^2 g_s(g_s'^2+1)+\lambda(g_s g_s'^2+6g_s+2g_s^2g_s''-3g_s^3)
\nonumber\\
&&+3g_s''g_s^2+6g_s-4g_s^3-g_s''\nonumber\\
&&C_1=-2g_s'(\lambda g_s^2+2)\nonumber\\
&&C_2=g_s(g_s^2-1)\nonumber
\ee

\begin{figure}[t]
\centerline{\includegraphics[width=8cm]{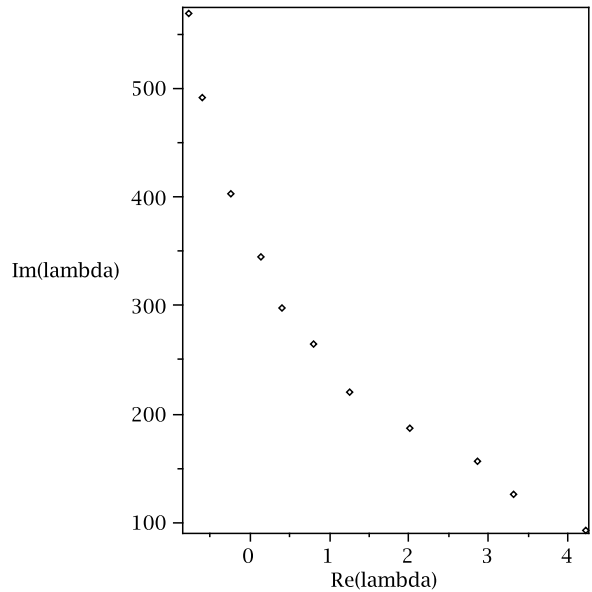}}
\caption{\small The evolution of eigenvalue $\lambda$ from Y=0.48 to 0.18
in the complex plane
} \label{eigenvalue}
\end{figure}

Due to the symmetry $y\leftrightarrow -y$ of the problem, we can solve it in 
the positive-$y$ region, with boundary condition $\psi(Y)=0$,$\psi'(0)=0$.
To solve this Schrodinger-like eqn, we use the iterative method. Starting
on one boundary with $\psi'(0)=0$,$\psi(0)=1$, the second condition only
affects the normalization of $\psi(y)$. With some initial value of $\lambda$,
we can obtain the $\psi(Y)$ from the EOM. then we variate the value such that
$\psi(Y)$ converge to 0. The resulting $\lambda$ gives the eigenvalue.
Without much difficulty, we found the following set of eigenvalue for
different $Y$, shown in Table.\ref{tab:eigen}. 
We also plot the eigenvalue $\lambda$ in the complex plane Fig.\ref{eigenvalue}.
The evolution trend of this set of eigenvalues
 suggests that the transition from stable to unstable 
occurs at $Y_m$ inside .22-.27 interval, which is way below
the critical value $Y\sim .5$ above which there were no scaling
solutions at all.
 This shows that we essentially lose the 
scaling solution to instability for $Y>Y_m$: we were not able to
tighten this limits any further.

\begin{table}
\caption{\label{tab:eigen} one set of eigenvalue for different rapidity}
\begin{tabular}{ccccc}
$\lambda(10^{-2})$& 4.2+94.8i& 3.3+126.7i& 2.8+157.5i& 2.0+188.5i\\
\hline
$Y$& 0.48& 0.45& 0.42& 0.40\\
\hline
$\lambda(10^{-2})$& 1.2+222.1i&0.78+265.7i& 0.38+299.5i& 0.12+346.4i\\
\hline  
$Y$& 0.37& 0.33& 0.30& 0.27\\
\hline
$\lambda(10^{-2})$& -0.27+404.2i& -0.63+492.9i& -0.80+569.8i\\
\hline
$Y$& 0.24& 0.21& 0.18\\
\end{tabular}
\end{table}

In summary, the scaling solution  exist
 only for sufficiently small rapidities $Y<Y_c\sim0.5$.
Furthermore, we were able to verify that it is classically unstable
already for $Y>Y_m\approx 1/4$. Therefore solutions other than the scaling one
is need for large rapidity, which is more important for our purpose.

\subsection{Falling strings: the non-scaling solutions}
\label{sec_strings_ads}
In this section we  study generic solutions outside
the scaling ansatz. But before we do so, let us explain qualitatively
why such solution must fail as the rapidity of the collision grows.
The scaling solution, in which $\tau$ and $y$ dependences
factorize, means that one tries to enforce a particular stable profile
to a string. But
as  the rapidity gap $2Y$ between the walls grows, 
we so-to-say try to build
wider and wider ``suspension bridge'' out of the string:
 it is going to break under its weight
at some point.

 We again use $z(\tau,y)=\tau g(\tau,y)$ and EOM (\ref{eq:nonscaling}).
 The boundary condition is
$g(\tau,y=\pm Y)=0$. Due to the symmetry of the problem, it is
sufficient to solve the dynamics of half of the string, with
initial condition $g(\tau,Y)=0$ and
$\frac{\partial g}{\partial y}(\tau,0)=0$.

However there are two potential problems in (\ref{eq:nonscaling}).
(i)the $y$ derivative diverges on the boundary.
(ii)the PDE is highly nonlinear and will show self-focusing
of energy at certain ``corners'', as we will see.
These make it difficult to obtain a well-behaved numerical solution
\footnote{Similar problems 
have been encountered by previous studies of jet quenching, and
another way to deal with them, proposed in Hertzog et al \protect\cite{jet}, 
 takes advantage
of the re-parametrization invariance to fine tune the performance of
PDE solver. }, 
and
 to improve the performance of Maple PDE solver  we
 used function $h(\tau,y)=g(\tau,y)^n$ as dynamical variable,
with properly chosen integer power $n$ so that the $y$ derivative
is finite on the boundary. 

\begin{figure}[t]
\centerline{\includegraphics[width=8cm]{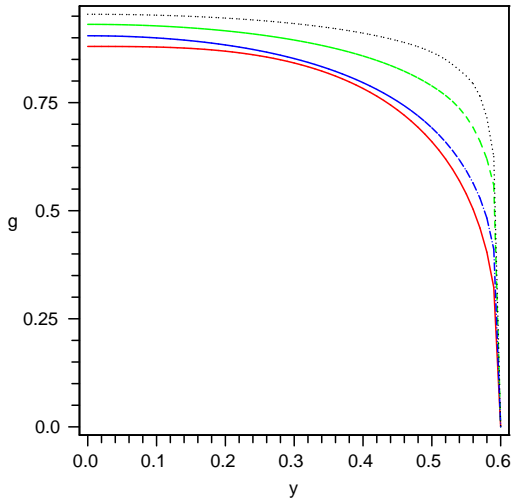}}
\caption{\small The dynamics of the string(half) $g(\tau,y)$ with $y=0.6$.
The profiles from the innermost to the outermost correspond
to $\tau=1$(solid red),$\tau=2$(dotted blue),$\tau=4$(dashed green),
$\tau=8$(dot-dashed black).
} \label{nonscaling}
\end{figure}

Fig.\ref{nonscaling} shows the dynamics of the string with $Y=0.6$. 
We start from
the initial condition $({g(1,y)\over 0.88})^3+({y\over Y})^3=1$
and $\frac{\partial g}{\partial \tau}(1,y)=0$. We choose the initial
time $\tau=1$ to avoid the singularity at $\tau=0$. $n=6$ is used in
solving the PDE.
As time grows, the string profile approach a rectangular shape with
sharper and sharper turn at the ``corners''. 
Based on the numerical solution, we infer that in the $\tau,y$
coordinates, any point of the string other than the boundary will
ultimately become free falling when time is sufficient large.
This can be supported by the following qualitative argument. Any tiny
piece of string experiences the AdS effective gravity and the drag 
from its neighbors. Since in the non-scaling solution, the whole string keeps
falling, it is natural to expect any point of the string approach the
speed of light asymptotically, end up with a rectangular profile.
Therefore, we conclude the edge of the profile is not important
asymptotically. It can be well approximated by a flat profile
in $y$, which will be studied in the next section.

\subsection{Falling strings and membrane in $AdS_5$}

  The falling string can be considered as a solution at the center
of the generic case considered above in the 
 large rapidity limit of the ends $Y\rightarrow \infty$.
 which makes $z$ $y$-independent. Ignoring
all derivatives over $y$ in the EOM above
one gets  an ODE problem with the following eqn:
\be
-2\,\tau+{ \dot z}\,z-{{ \dot z}}^{3}z+\tau\,{\ddot z}\,z+2\,{{\dot z}}^
{2}\tau=0
\ee
which is similar but not identical to that of a falling massive
object (\ref{eqn-stone}): 
the difference comes from dimensionality of the
object: $1/z^2$ in the action
(instead of $1/z$), because the string action is a 2-dimensional
integral. 
It is now explicitly depending on $\tau$: there is no integral of
motion but one can
 straightforwardly solve the EOM for
different initial conditions numerically.
 We found at large $\tau$, $g$ tends to 1.
Therefore we show in this extreme case that the asymptotic solution
is again $z\sim\tau$

Summarizing the falling of all string objects, they have a universal
asymptotic behavior $z\sim\tau$. Therefore we may model the falling 
particles/open strings by a membrane, which is 
made of multiple strings and is flat in $x_2$, $x_3$ and $y$ coordinates

 The coefficient in its
 DBI action, the membrane tension,
is now proportional to the density of charges in the colliding walls,
and thus can be very large. This fact would mean that the membrane
should eventually be considered heavy enough, so that
its weight would affect the metric itself.
Since in this work we would not attempt to solve this problem yet, 
we treat the membrane as a test body falling in external AdS
metric. In this case the value of its tension does not matter,
and the action is very similar to Nambu-Goto string action
except of the different power of $z$ (now $1/z^4$)
 
\be
 S\sim \int \tau d\tau dydx_2dx_3 {\sqrt{1-({\partial z\over \partial \tau})^2}
 \over z^4}  
 \ee

We parametrize the membrane with $\tau$,$y$,$x_2$,$x_3$, and
z-coordinate is a function of $\tau$ only, $z=z(\tau)$.
 The EOM is readily obtained, it is similar to
 the $y$-independent string case (coefficients 2 change to 4
in two terms):
\be
{\dot z}z-{\dot z}^3z+4\tau{\dot z}^2+\tau{\ddot z}z-4\tau=0
\ee
Its asymptotic solution is again $z\sim\tau$.

\section{Near-horizon ``braking''}
\subsection{Stretching black holes}
\label{sec_jp}
  
 The JP solution we will now discuss
addresses the first case, d=1. 
The main feature of the JP solution is that these two variables enter
 the metric via one specific combination
\be v={z \over \tau^\gamma  } \ee
which simplifies Einstein's eqns and leads to a solution.
JP have found that only for one particular power $\gamma=1/3$
there is no singularity at the horizon in one of the invariants --
the square of the 4-index Riemann curvature, and argued that thus this
solution should be preferred on this ground.
 
 However it is not clear what  the physical meaning
 and  significance of this singularity may be, in general.
Furthermore, in the ``membrane scenario'' proposed in this work
the JP-like metric only extends from the AdS boundary till the
falling membrane, while the would-be singularity is in the second
 domain, where this solution is not supposed to be used at all.
It is, so to say, a ``mirage behind the mirror'',
singular or not does not matter.

  There is another reason why this particular power should be selected:
only in $\gamma=1/3$  case such that
the total area of the horizon (3d object normal to time and z)
is  {\em time independent}: the factor $\tau$ (from
 stretching
$y_1$) is canceled by the factor $1/z^3$ from contracting $z$. 
Thus, this stretching solution is area-preserving, and thus
potentially dual to the entropy-conserving adiabatically expanding
 fireball.

The specific form of the JP metric is
\be
ds^2=-\frac{(1-v^4{e_0\over3})^2}{(1+v^4{e_0\over3})}\frac{d\tau^2}{z^2}+
(1+v^4{e_0\over3})\frac{\tau^2 dy^2+dx_\perp^2}{z^2}+\frac{dz^2}{z^2}
\ee
The horizon determined from $g_{\tau\tau}(v)=0$ is
at $v_h=({3\over e_0})^{1/4}$,  thus it is moving
away from $z=0$ (the AdS boundary) as needed.
The 4-th power of $v$ is related to the fact that 
 its expansion near $z=0$ to the 4-th order is responsible for
the stress tensor as observed on the boundary, which was tuned to 
correspond to the  Bjorken
boost invariant solution of ideal hydrodynamics \cite{Bj}:
the starting point for JP.

This metric provides an asymptotic (large $\tau$) 
solution to the Einstein eqns
\be R_{\mu\nu}-(R/2)g_{\mu\nu}-6g_{\mu\nu}=\kappa T_{\mu\nu}\ee
  After this metric is
 substituted to the l.h.s. one finds that all terms of the 
``natural'' order of magnitude
$O(\tau^{-2/3})$ cancel out, with only the higher 
order terms remaining. More specifically, we found that only the
terms $T_{\mu\nu}\sim 1/\tau^2$ are present, with rather compact 
expressions such as
  \be \tau^2 T_{\tau\tau}=-{4v\over (3+v)^2} \ee
\be \tau^2 T_{zz}=-{4v^2 \over (3+v)(v-3)^2} \ee
\be \tau^2 T_{yy} = (-4/9) {v (4v^2-15v-63) \over (v-3)^3}\ee
Please note that those terms are not only subleading
at large $\tau$ but also are much simpler than
all the terms which had canceled out. Also note that
there is a significant singularity at the horizon
($v=3$ in these units) in this stress tensor, which is
again irrelevant because this metric is not supposed to be
used there.

\subsection{Objects approaching the horizon}
\label{sec_objects_JP}
Before we discuss the JP metric, let us remind the reader how
this approach works in the usual black holes with the Schwartzschild
metric: it will be needed to emphasize the
difference between them.

Massless particle falling radially in the
  Schwartzschild metric satisfies the $ds^2=0$ eqn, which is
\be ({dr \over dt})^2=(1-{r_h\over r})^2\ee
leading to exponentially fast ``freezeout'', 
\be (r-r_h)\sim exp(-t/r_h) \ee
The same is also true for other objects, of course.

We use the following rescaled coordinates:

\be
z\rightarrow c\,z, \tau\rightarrow c\,\tau, y\rightarrow y, x_{\perp}\rightarrow c\,x_{\perp} \nonumber
\ee
with $c=({\frac{3}{e_0}})^\frac{3}{8}$. The resultant metric is
\be
\label{eq:jp_c}
ds^2=-\frac{\left(1-{\frac {{z}^{4}}{{\tau}^{4/3}}} \right)^{2}}
{1+{\frac {{z}^{4}}{{\tau}^{4/3}}}}\frac{d{\tau}^2}{z^2}
+ \left( 1+{\frac {{z}^{4}}{{\tau}^{4/3}}} \right)  
\frac{{\tau}^{2} dy^{2}+dx_{\perp}^{2}}{z^2}+\frac{dz^2}{z^2}
\ee
The
massless particle moves according
to $ds^2=0$, which in JP metric is
\be
\frac{dz}{d\tau}=\frac{1-{\frac {{z}^{4}}{{\tau}^{4/3}}}}
{\sqrt{1+{\frac {{z}^{4}}{{\tau}^{4/3}}}}}
\ee
We have assumed that the particle always starts from outside the horizon: $z<\tau^\frac{1}{3}$
This EOM is solved numerically for different initial conditions.(From here on, we always use
$\tau=10$ as initial time for numerical solution, since the metric (\ref{eq:jp_c}) is valid
asymptoticly $\tau>>1$)

\begin{figure}[t]
\centerline{\includegraphics[width=8cm]{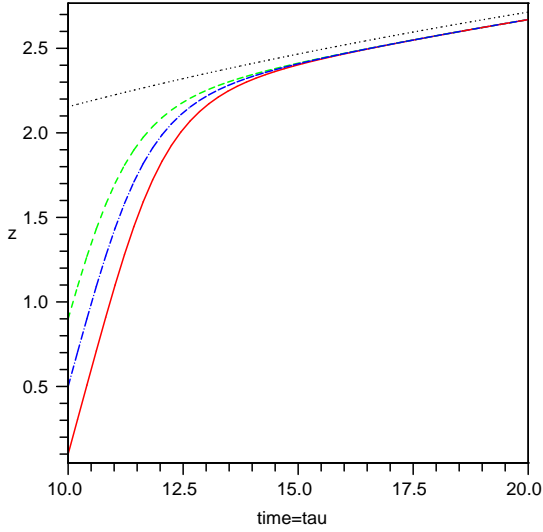}}
\caption{\small trajectories of massless particles, with
initial z coordinates: z(10)=0.1(solid red) z(10)=0.5(dash-dotted blue) 
z(10)=0.9(dashed green)
The horizon is also plotted(dotted black) for comparison. The trajectories of the massless particles
approach each other asymptotically, but does not seem to
 approach the moving horizon.
} \label{fig_falling_massless}
\end{figure}

To obtain the analytical form of the asymptotic behavior, we define:
\be u=\frac{z^4}{\tau^{\frac{4}{3}}}\ee
and the EOM becomes
\be
{\frac {1-u}{\sqrt {1+u}}}=\frac{1}{4}\,{\frac {\dot u\,{\tau}^{1/3}}{{
u}^{3/4}}}+\frac{1}{3}\,{\frac {{u}^{1/4}}{{\tau}^{2/3}}}
\ee

Note $u\rightarrow 1$ as $\tau\rightarrow\infty$. Assuming the second term dominates
the first term on the RHS, we obtain the asymptotic
form $u=1-\frac{\sqrt{2}}{3}\,\tau^{2/3}$, which confirms our assumption.
In terms of $z$ and $\tau$, we have:

\be
\label{eq:massless}
z=\tau^{1/3}\,\left(1-\frac{1}{6\sqrt{2}}\,\tau^{-2/3} \right)
\ee

For massive particle, the action is given by $S=m\int{ds}$. Similarly
we focus on the case that particle moves in a trajectory with constant $y$ and $x_{\perp}$:
EOM follows from variation on action.
%
Let $z=\tau^{1/3}\,f$, then the function $f$ needs to satisfy the
following eqn:

\be
\label{eq:massive_f}
&&-27{\tau}^{2}{f}^{16}{{\dot f}}^{2}-6\tau{f}^{17}{\dot f}+18{\tau}^{2}{f}^{8}{{\dot f}}^{2}
-108{\tau}^{2}{f}^{12}{{\dot f}}^{2}-6{f}^{14} \nonumber\\
&&+4{f}^{10}+54\tau{f}^{5}{\dot f}
-54\tau{f}^{13}{\dot f}+12\tau f{\dot f}+108{\tau}^{2}{f}^{4}{{\dot f}}^{2}\nonumber\\
&&-6\tau{f}^{9}{\dot f}+6{f}^{6}-3{f}^{18}+9{\tau}^{2}{f}^{17}{\ddot f}
+9{\tau}^{2}f{\ddot f}-9{\tau}^{4/3}\nonumber\\
&&-18{\tau}^{2}{f}^{9}{\ddot f}
-126{\tau}^{4/3}{f}^{12}+9{\tau}^{4/3}{f}^{20}+27{\tau}^{4/3}{f}^{16}\nonumber\\
&&-27{\tau}^{4/3}{f}^{4}+126{\tau}^{4/3}{f}^{8}-{f}^{2}
+9{\tau}^{2}{{\dot f}}^{2}=0
\ee

It is again solved numerically, with initial conditions satisfying
$z_0<\tau_0^{1/3}$ and $\dot z(\tau_0)<\frac{1-{\frac {{z_0}^{4}}{{\tau_0}^{4/3}}}}
{\sqrt{1+{\frac {{z_0}^{4}}{{\tau_0}^{4/3}}}}}$.
Note that free falling massive object will move with speed of light 
asymptotically. We expect (\ref{eq:massless}) to be the asymptotic
solution. By plugging (\ref{eq:massless}) in (\ref{eq:massive_f}), we
get the RHS: ${8\over 3}\tau^{-4/3}$, which tends to zero as $\tau$ grows

\begin{figure}[t]
\centerline{\includegraphics[width=8cm]{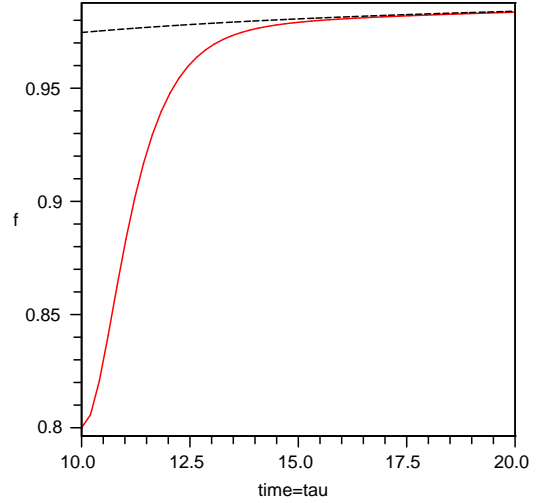}}
\caption{\small trajectory of massive particles starting
with $f=0.8$ and $\dot f=0$.(solid red) at $\tau=10$.
The trajectory is indistinguishable from the asymptotic solution(dashed black)
 at $\tau\sim 15$
} \label{stone_jp}
\end{figure}

Furthermore, we compare numerical solution with the asymptotic solution 
in Fig.\ref{stone_jp}. The two solutions agree well at large $\tau$.
This confirms (\ref{eq:massless}) is the correct asymptotic solution.

To study the falling string, we first parameterize the string by $z=z(\tau,y)$.
Instead of solving it this form. We recall our
experience with non-scaling solution in AdS space.
At large enough $\tau$, the edge of the string
will be less important, with most part of the string falling freely. 
Therefore we ignore the
y dependence of z: $z=z(\tau)$

Defining $f=\frac{z}{\tau^{1/3}}$, The EOM follows straightforwardly from the Nambu-Goto action with the metric (\ref{eq:jp_c}). 
It is a quite lengthy expression, which we choose not to show here.

\begin{figure}[t]
\centerline{\includegraphics[width=8cm]{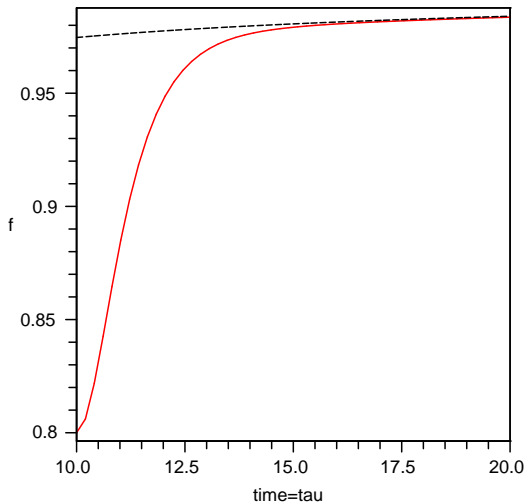}}
\caption{\small trajectory of string with initial condition $f=0.8$ and
 $\dot f=0$(solid red) at $\tau=10$.
The trajectory is indistinguishable from the asymptotic solution(dashed black)
 at $\tau\sim 15$
} \label{string_jp}
\end{figure}

 
We expect the same asymptotic
solution (\ref{eq:massless}). By plugging (\ref{eq:massless}) in
the EOM, we get the RHS: $-\frac{95\sqrt{2}}{12}\tau^{-2/3}$, which
 tends to zero as $\tau$ grows.
Fig.\ref{string_jp} compares numerical solution with the asymptotic solution,
which confirms it is the correct asymptotic solution.

Now we proceed to our final case, a $membrane$ falling in JP metric.
Let $z(\tau,y,x_2,x_3)=\tau^{1/3} f(\tau)$:
the EOM is again quite lengthy and not shown here.


We have solved it with a number of initial conditions
and found that all extra terms are subleading
near horizon, so this EOM gives the same asymptotic
solution as the other cases,
namely $f= 1- ({1\over {6\sqrt{2}}})\tau^{-2/3}$.

\begin{figure}[t]
\centerline{\includegraphics[width=8cm]{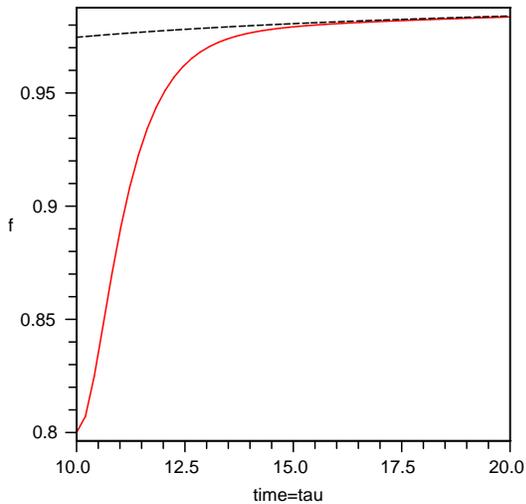}}
\caption{\small trajectory of membrane with initial condition $f=0.8$ and
 $\dot f=0$(solid red) at $\tau=10$.
The trajectory is indistinguishable from the asymptotic solution(dashed black)
 at $\tau\sim 15$
} \label{brane_jp}
\end{figure}

The numerical solutions are displayed in Fig.\ref{brane_jp}, which confirm
the asymptotic solution.

We found that in all cases studied --
 massless and massive particles,  string and membranes -- 
their late time behavior can be
approximated by the same asymptotic solution 

\be (z-z_h(\tau))\sim [-{1\over {6\sqrt{2}}}\tau^{-1/3}+...] \ee

\section{Summary}
\label{sec_summary}
  This is the first paper of the series, devoted to
quantitative formulation of the ``gravity dual''
to high energy collisions of macroscopically large bodies (heavy
ions). In it we have formulated the setting in which
the problem is simplified sufficiently to be solvable.

  Its central idea is that various ``debris'' from a collisions, 
in form of massless and massive particles or ``stretching'' open
strings, all fall toward the AdS center. Although qualitatively
such falling may look quite similar, the equations of motion and solutions
are different for different objects.
The main result of this work is a systematic 
 demonstration of this statement in detail, both
for initial time (when 
the underlying metric is supposed to be close to AdS) and
at the late times (when the metric is close to JP solution).
As we will see in subsequent papers later, small differences
in ``falling'' leads to quite different ``holograms'' in form of
stress tensor at the boundary.

One possible solution can be to unify all such ``debris''
 as a single massive ``membrane'', falling under its own weight.
As shown first by
Israel \cite{Israel} long ago, in such case one can greatly
simplify the gravitational aspect of the problem, using two different
 solutions of the sourceless Einstein equations
inside both  space-time domains, appropriately matched at the
hypersurface made by the world-volume of the
membrane. Two solutions are subject to ``junction conditions''
providing new EOM for the membrane itself. We will discuss those
issues elsewhere.

Let us now point out few more specific results of this work.
In the study of
longitudinally stretched strings we have found that ``scaling''
solutions used previously for determination of ``kink'''s anomalous
dimensions are not at all adequate  in Minkowski time.
We found that while for wall rapidity $Y>Y_{max}\approx 1/2$ these
solutions are absent, and there are two of them for smaller $Y$.
We further studied stability of the solutions and have proven
that at least for $Y>Y_c\sim 1/4$ they indeed are unstable.

Our main finding for generic non-scaling solutions
(which come from numerical solutions of PDEs)  
is that while at small velocity of stretching
there is the so called scaling solution, generically at high
stretching one gets instead asymptotic approach to a ``rectangular''
 solution, consisting basically of two near-vertical strings and freely
 falling horizontal part.

  Another result which was not expected is that 
although all types of objects
-- massless and massive particles as well as open strings and
membranes -- approach the
 JP  horizon in the same universal way. 
Unlike in the textbook case of the
Schwartzschild metric, this approach does 
 not happen exponentially but only as
a power $\tau^{-2/3}$ of time. 
Note that this power is the same as appears in subleading
terms,  ignored by JP at late time. It remains a challenge to
find an appropriate vacuum solution to Einstein equation
complementing the late-time JP metric.

\noindent{\large \bf Acknowledgments} \vskip .35cm We thank
Ismail Zahed and San-Jing Sin for multiple discussions. Our
 work was partially
supported by the US-DOE grants DE-FG02-88ER40388 and
DE-FG03-97ER4014.s

\vskip 1cm

\end{document}